# CarPed – A hybrid and macroscopic traffic and pedestrian simulator

Daniel H. Biedermann[1] und Peter M. Kielar[1] und Quirin Aumann[1] und Carlos M. Osorio[1] und Celeste T. W. Lai[1]

[1]Lehrstuhl für Computergestützte Modellierung und Simulation · Technische Universität München · Arcisstraße 21 · 80333 München
daniel.biedermann@tum.de

## Zusammenfassung

Dichte Personenströme sind ein bekanntes Gefährdungsrisiko für öffentliche Großveranstaltungen. Makroskopische Personenstrommodelle, die häufig auf fluid-dynamischen Berechnungen basieren, können mit geringem Rechenaufwand simuliert werden. Aus diesem Grund werden sie oft für die Simulation großer Menschenmengen verwendet (COLUMBO & ROSINI 2005). Ähnliche Ansätze existieren im Forschungsbereich der Verkehrssimulation (LIGHTHILL & WHITHAM 1955). Eine gekoppelte makroskopische Simulation von Fahrzeugen und Fußgängern ist für eine allumfassende Verkehrsüberwachung und Planung sehr hilfreich. Aus diesem Grund haben wir ein Hybridmodell entwickelt, das Netzwerke für Fahrzeuge und Fußgänger bereitstellt. Dieses gekoppelte Modell unterstützt die gleichzeitige Simulation von Fahrzeug- und Personenverkehr.

## Abstract

Dense human flow has been a concern for the safety of public events for a long time. Macroscopic pedestrian models, which are mainly based on fluid dynamics, are often used to simulate huge crowds due to their low computational costs (COLUMBO & ROSINI 2005). Similar approaches are used in the field of traffic simulations (LIGHTHILL & WHITHAM 1955). A combined macroscopic simulation of vehicles and pedestrians is extremely helpful for all-encompassing traffic control. Therefore, we developed a hybrid model that contains networks for vehicular traffic and human flow. This comprehensive model supports concurrent multi-modal simulations of traffic and pedestrians.

## 1 Introduction

### 1.1 Motivation

Every year, about 2000 people die due to the movement of large human crowds (HUGHES 2003). The growing eventisation of social and public life in the 21st century (HITZLER 2010) increases the probability and frequency of organized events attended by thousands or millions of people. In the past, fatal accidents have occurred repeatedly at such large events. Examples include the catastrophe in the Heysel stadium in 1985 (LEWIS 1989) with 39 fatalities or the Loveparade disaster in 2010 with 21 deaths (HITZLER ET AL. 2011). Professional crowd control increases the safety of public events and lowers the probability of fatal accidents in the context of large human flows. Pedestrian dynamics simulations support successful crowd management by helping organizers of public events to foresee and prevent dangerous situations. Macroscopic pedestrian dynamic models like the Hartmann-Sivers model (HARTMANN & SIVERS 2013) are most suitable for simulating large numbers of people (BIEDERMANN ET AL. 2014A). The same macroscopic model can be applied to simulate the motion of cars. By combining human and vehicular flow into one model, we can achieve an all-embracing simulation model, which covers crowd and traffic control at the same time. Due to the macroscopic nature of these simulation models, we can simulate very large and dense scenarios faster than real time.

## 1.2   Current Hybrid Modeling

Two different kinds of hybrid modeling exist in the field of traffic and crowd control. The first type combines traffic or pedestrian dynamcis models at different scales in order to reduce the overall computational effort. This means that simulations in the hazardous areas of the scenario (places with high densities or small bottlenecks) are calculated with very detailed but computationally costly models. At the same time, less hazardous areas of the scenario can be simulated by less detailed but more cost-effective models. This provides us precious results in the interesting parts of the scenario with a reduced total computational cost. TOLBA ET AL. (2005) as well as MCCRE & MOUTARI (2010) developed such hybrid models in the field of vehicular flow. Similar models were developed for pedestrian dynamics. A wide range of hybrid modeling exists; from the coupling of very specific pedestrian simulation models (e.g. NGUYEN ET AL. 2012) to generic coupling frameworks (BIEDERMANN ET AL. 2014B). An overview of the current state of the art can be found by IJAZ ET AL. (2015).

The second type of hybrid modeling does not combine different scales, but different kinds of simulations. GALEA ET AL. (2008) describe the coupling of pedestrian dynamics and fire simulations and GÖTTLICH ET AL. (2011) combine an evacuation simulation with the spread of hazardous gases. Our approach is the connection of vehicular traffic with human flow. Some studies have already been carried out in this field. PRETTO ET AL. (2011) coupled agent-based representations of pedestrians and cars. A macroscopic approach was developed by BORSCHE ET AL. (2014). They used the classic Lighthill-Whitham-Richards model (LIGHTHILL & WHITHAM 1955; RICHARDS 1956) for the vehicular flow and the model from HUGHES (2002) for the macroscopic simulation of pedestrians. Their work is important progress in the field of hybrid macroscopic modeling. However, they considered vehicles and humans as two interacting, but separate flows. In reality, cars and pedestrians are not completely separate: drivers exit their cars after finishing driving and become pedestrians and pedestrians may enter their vehicles to become drivers. Our approach is a hybrid model, which is capable of converting humans and vehicles into each other to create a realistic simulation model.

# 2   The hybrid simulation model CarPed

## 2.1   The Hybrid approach

It is important to use a single network that combines pedestrian and traffic simulations to enhance the safety of traveling to public events. Therefore, we introduce an interface which links human and vehicular flow. The movement behavior of cars and pedestrians is based on the macroscopic Hartmann and Sivers model (HARTMANN & SIVERS 2013) and is calculated on a discrete network. Roads for cars and walking paths for pedestrians are represented by the edges of this network. Its nodes represent the links between the different edges and can be considered crossings. Specifically, the nodes have different flow rates. There are special nodes, which represent parking lots, and serve as a connection between roads and walkways. Visitors of the event enter and leave their cars there. To put it another way, the simulated subjects get transformed from vehicles into pedestrians and vice versa. Therefore, the parking lots serve as a connection of the traffic and pedestrian networks.

In Figure 1, an exemplary network is shown. The darker nodes and edges represent streets and their respective intersections, whereas the pedestrians' walkways and intersections are represented in a lighter gray. The transformation of cars and pedestrians is carried out at special parking lot nodes. Green numbered entry nodes represent sources, which are the starting points of the visitors (e.g. their hometown). The red marked exit node is the final destination of the visitors (e.g. a large public event), where they are deleted from the network.

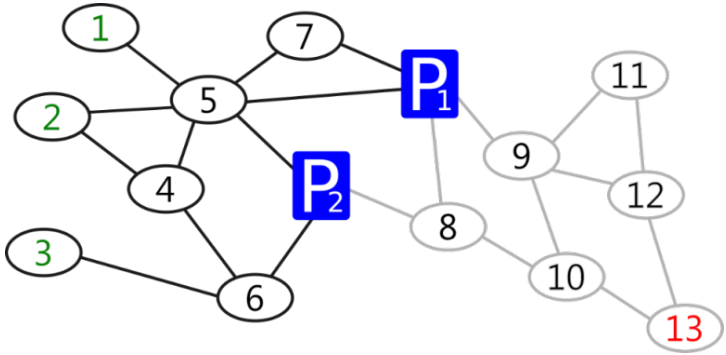

**Figure 1:** An exemplary network consisting of nodes and edges

We implemented the Hartmann and Sivers model for pedestrian simulation and extended it to traffic simulation by using different model parameters. The extension was done by applying different maximum capacities to the edges and nodes, using different free flow velocities and other model constants for the simulated subjects. We used data obtained from WEIDMANN (1993) for the pedestrians and data from WACHS ET AL. (2000) for the vehicle simulation. Additionally, we conducted a large field study to receive sufficient data for the transformation process (see Section 3).

## 2.2   The Hartmann-and-Sivers Model

The Hartmann and Sivers model adopts a structured continuum model on a macroscopic level. The approach basically relies on fundamental diagrams – the relation between fluxes and local densities – as well as the explicit consideration of individual velocities; thus showing better agreement with microscopic pedestrian models (HARTMANN AND SIVERS 2013). The free flow velocity $v_{ff}$ is an individual property of each pedestrian / vehicle, i.e. a macroscopic, time independent state variable attached to each individual. The free flow velocity describes the highest velocity an individual can reach under optimal circumstances.

This property is distributed normally among pedestrians or cars. Thus for modelling human or vehicular flow, the following equation holds

$$\frac{\partial \rho(v_{ff};x,t)}{\partial t} + \frac{\partial}{\partial x}\left(v(v_{ff};\rho(x,t))\rho(v_{ff};x,t)\right) = f(v_{ff};x,t) \qquad (1)$$

The total density of subjects is given by the sum of densities for all free flow velocities.

$$\rho(x,t) = \int_0^{v_{ff}^{max}} \rho(v;x,t)dv \qquad (2)$$

The fundamental relation between velocity and density can be adopted according to:

$$v(v_{ff};\rho) = v_{ff}(1 - \exp(\gamma(1/\rho - 1/\rho_{max}))) \qquad (3)$$

The maximum density $\rho_{max}$ and the factor $\gamma$ are unique parameters for cars and pedestrians. The solvation of the differential equations for pedestrians and cars was done by an upwind-downwind finite volume scheme (HARTMANN & SIVERS 2013).

## 2.3  Routing Behavior of Pedestrian and Cars

In their original paper, Hartmann and Sivers used fixed probability values to distribute the pedestrians at each intersection. These values were obtained by comparing the capacities of the adjacent edges and did not take the current density of those edges into account. It is possible that the majority of a flow gets directed into an edge which is already relatively full, and the required time for the flow to pass the system could increase in an unrealistic way. Therefore, we extend this approach and introduce a routing algorithm which determines the possible edges the pedestrians or cars will be allocated at.

To find the fastest path through the system, we use the classical Dijkstra's algorithm (DIJKSTRA 1959). This algorithm weighs all edges of the graph to compute the shortest way through the system. We used the Dijkstra's algorithm to calculate the route with the shortest travel time. Therefore, we used the length and the current velocity $v(v_{ff};\rho)$ as weighting factors:

$$weight = \frac{length}{v(v_{ff};\rho)} \qquad (4)$$

With increasing density the velocity $v(v_{ff};\rho)$ decreases according to Equation (2). If the density of an edge reaches the maximum density $\rho_{max}$, the edge is considered as closed. Closed edges are ignored for the calculation of the routing algorithm.

## 2.4  Transformation between the two simulation models

Cars get transformed into pedestrians and vice versa in the parking lot nodes. Typically, for ordinary cars, one to five people get out of each vehicle, with a prescribed random distribution. The distribution for the transformation was gathered from field data (see Section 3). It is also possible to transform incoming pedestrians into cars, using the inverse process. The transformed pedestrians and cars receive their individual parameters, like their free flow velocity or their maximum density, according to WEIDMANN (1993) respectively WACHS ET AL. (2000).

## 3 Visitor distribution of incoming cars

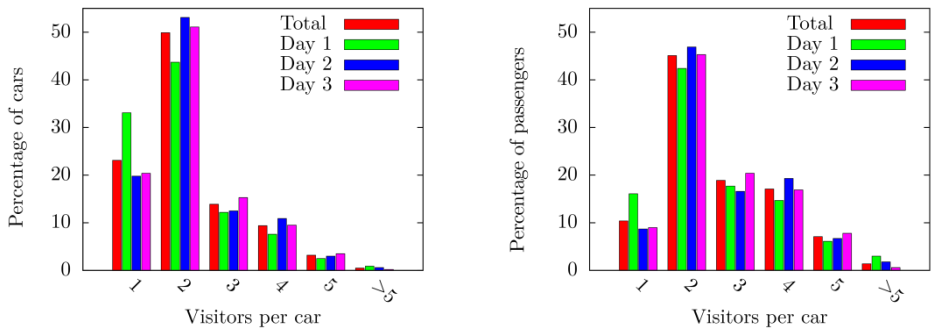

**Figure 2:** The distribution according to the total number of cars counted and the distribution according to the total number of passengers.

Pedestrians get transformed into cars and vice versa on special parking nodes in the graph based scenario. To model a realistic transformation of pedestrians and cars, it is necessary to know the distribution of passengers per car. This means, that we have to know how many visitors are normally transported by one car. Therefore, we conducted an extensive field study to obtain the lacking data. We studied a large music festival in Munich over three consecutive days and counted the amount of visitors per car. We surveyed a total number of 1960 cars (for details see Table 1), which carried an average of 2.21 persons. Over 70% of the vehicles were occupied by one or two passengers. A negligible amount of cars transported more than five people, and the maximum observed was eight passengers per car. The observed distributions from all three days can be seen in Figure 2. This data was used as a configuration for the transformation process of the hybrid model.

| Date \ Visitors | 1 | 2 | 3 | 4 | 5 | >5 | Total |
|---|---|---|---|---|---|---|---|
| 29.05.2015 | 144 | 190 | 53 | 33 | 11 | 4 | 435 |
| 30.05.2015 | 98 | 263 | 62 | 54 | 15 | 3 | 495 |
| 31.05.2015 | 210 | 526 | 158 | 98 | 36 | 2 | 1030 |
| Total | 452 | 979 | 273 | 185 | 62 | 9 | 1960 |

**Table 1:** Number of visitors per car on the music festival "Rockavaria"

## 4 The CarPed-Toolbox

### 4.1 Structure and usability

In cooperation with professional and experienced event managers, the software toolbox CarPed was developed as a proof of concept. Great attention was payed towards usability. Therefore, we implemented an intuitive and easy to use graphical user interface (GUI).

The GUI is divided into three major parts, namely the network input, the model and solving options and the result window. The network, consisting of nodes and edges, can be easily entered through the network input. All network depending properties, like the length or the capacity of an edge, can be determined immediately after the input procedure. By using the model and solving options window, the user can determine which computational method should be used and how the data should be processed. Finally, the computed results can be accessed and visualized in the viewer. As all computed data is stored, the density distribution for each time-step can be looked at by the user. Different coloring of the edges shows the current densities. Alternatively, the user can click on an edge or node to get more detailed information about this object like the amount of pedestrians or cars on it. An exemplary screenshot of the GUI can be seen in Figure 3.

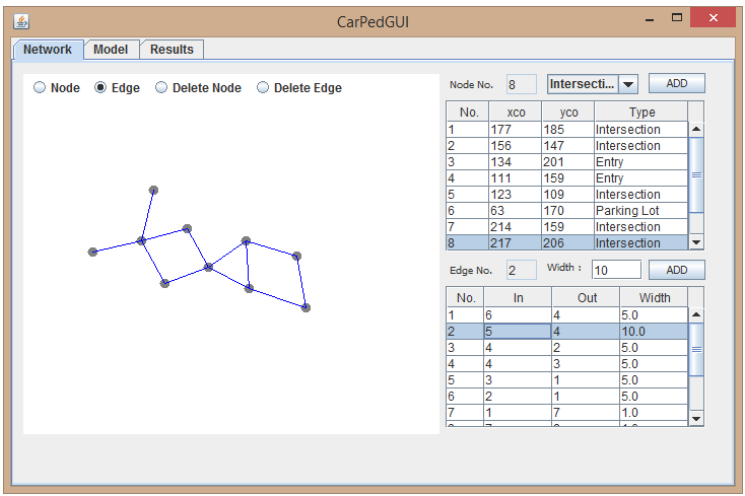

**Figure 3:** Representation of an exemplary simulation scenario in CarPed

### 4.2   Workflow

The network can be entered by clicking on the designated area in the GUI. Special nodes, like entry and exit nodes or parking lots, can be defined by the user. Additionally, the nodes can be linked to incoming or outgoing edges. If the input to the system is finished, a checking routine searches for errors and discrepancies. For each algorithm, the user can individually modify the settings. After everything is set up, the solver begins its calculations.

At the beginning of the simulation, the network model reads the information of nodes and edges from the database previously created. It constructs a network from the infrastructure and passes the entire network to the numerical solver. Together with the received input data, the numerical solver starts generating vehicles and pedestrians in the associated entry nodes. It determines the initial density for each edge that is connected to the entry node and computes the density distribution. For each time-step, the results are passed from the numerical solver and stored in the network model. If a subject enters a node, it gets removed from this incoming edge and is put into an outgoing edge according to the route calculated by the Dijkstra algorithm. Additionally, the subject gets transformed if the visited node is a parking lot. The computations are repeated until all pedestrians and cars have reached an exit node and are dismissed from the system.

During the result analysis, the GUI displays two visions: a parameter set and an animation of density distributions. All statistical information is provided by the toolbox and can be easily accessed.

## 5      Conclusion and Outlook

This paper describes the dedicated hybrid and macroscopic simulation approach CarPed. This approach uses the macroscopic Hartmann and Sivers model on a network of nodes and edges. It is able to simulate human and vehicular flow in one hybrid model. The transformation between pedestrians and cars is carried out at special nodes, which represent parking lots.

The tool is thought to play a fundamental rule in the planning phase of massive public events. Organizers can use it to identify pathways and roads, which have a higher risk of congestions. Therefore, the organizers can develop strategies to avoid dangerous situations. In future studies, the current approach will be extended to a hybrid model for all-encompassing traffic control. This extended hybrid model will include more traffic subjects, like trains or ships to receive a global and universal traffic simulator.

Additionally, the CarPed simulator needs to be validated by field data. Therefore, we observed a public music festival over two consecutive years (BIEDERMANN ET AL. 2015). The first results are promising, but need further investigations.

## Acknowledgements

We would like to thank Isabella von Sivers for her support and long conversations according to this topic. Additionally, we would like to thank Andrea Mayer, Andreas Riedl and our other student assistants for their help to collect sufficient data.

This work is supported by the Federal Ministry for Education and Research (Bundesministerium für Bildung und Forschung, BMBF), project MultikOSi, under grant FKZ 3N12823.